\documentclass[aps,prb,twocolumn,superscriptaddress]{revtex4}
\usepackage{graphicx}
\usepackage{bm}
\usepackage{amsmath}
\usepackage{amsfonts}
\usepackage{amssymb}
\usepackage{epsfig}

\newcommand{\mume}{\mu\textrm{m}}

\begin{document}
\title{Quantum Monte Carlo study of ring-shaped polariton parametric luminescence in a semiconductor microcavity}

 \author{A. Verger}
 \affiliation{Laboratoire Pierre Aigrain,
Ecole Normale  Sup\'erieure, 24, rue Lhomond, 75005 Paris, France}
\author{I. Carusotto}
\affiliation{BEC-CNR-INFM and Dipartimento di Fisica, Universit\`a
di Trento, I-38050 Povo, Italy}
\author{C. Ciuti}
\email[E-mail: ]{cristiano.ciuti@univ-paris-diderot.fr}
\affiliation{Laboratoire Mat\'eriaux et Ph\'enom\`enes Quantiques,
UMR 7162, Universit\'e Paris 7, 75251 Paris, France}

\begin{abstract}
We present a quantum Monte Carlo study of the quantum correlations
in the parametric luminescence from semiconductor microcavities
 in the strong exciton-photon coupling regime.
As already demonstrated in recent experiments, a ring-shaped
emission is obtained
  by applying two identical pump beams with opposite in-plane wavevectors, providing
   symmetrical signal and idler beams with opposite in-plane wavevectors on the ring.
 We study the squeezing of the
     signal-idler difference noise across the parametric instability threshold,
      accounting for the radiative and non-radiative losses, multiple scattering
       and static disorder. We compare the results of the complete multimode Monte
        Carlo simulations with a simplified linearized quantum Langevin analytical model.

\end{abstract}
\date{\today} \maketitle

In the last years, semiconductor microcavities in the strong
exciton-photon coupling
regime\cite{review_Cristiano,review_Deveaud} have been attracting
a considerable deal of interest because of their remarkable
nonlinear parametric
interactions~\cite{Baumberg,Nature_Diederichs,Savvidis_00,Gregor_01,Gregor_02}:
taking advantage of a
triply-resonant condition, ultra-low parametric oscillation
thresholds have been observed in geometries which look very
promising in view of
applications.
Very recently, experimental and theoretical investigations are
starting to address the genuine quantum optical properties
 of the polariton parametric emission
\cite{entanglement_Ciuti,Iac_Crist_1,Carusotto_QMC,Karr_squeeze,
Langbein_complementarity,degen_Baas,twin_Karr}.
 The signal-idler pairs generated by the coherent scattering of two
 pump polaritons are expected to have non-classical properties, such
 as entanglement and two-mode squeezing, which are interesting
 e.g. for quantum teleportation.
The main limitation of the original non-degenerate parametric
scheme where the cavity was pumped by a single incident beam at a
finite 'magic'
angle\cite{Karr_squeeze,degen_Baas,Savvidis_multiple}, was the
strong intensity asymmetry between the signal and idler photon
emission. This signal-idler asymmetry is in fact strongly
detrimental in view of the observation of significant extra-cavity
quantum correlations to be used for continuous variable
experiments.

 This difficulty has been overcome in recent
experiments\cite{Romanelli_Leyder} by using a pair of identical
pump beams with small and opposite in-plane wavevectors. In this
degenerate parametric scheme, a pair of perfectly symmetric signal
and idler beams are emitted at the same frequency and with
opposite wavevectors.
For symmetry reasons, the momentum-space parametric luminescence
pattern is in this case a ring, with approximately the same radius as
the pump wavevector. Interestingly, this kind of ring-shaped polariton
          parametric luminescence can be obtained also with a single
          pump at normal incidence (zero in-plane wavevector) on a
          multiple microcavity with multiple photonic
          branches\cite{Diederichs_Leyder}.
In order to quantify the performances of this system as a source of
correlated photons, it is then important to characterize the
          robustness of the quantum correlations in the parametric
          luminescence against competing effects such as
          radiative           and non-radiative losses as well as
          multimode competition
          and multiple scattering processes.
Given the unavoidable imperfections of any solid-state system, it is
also crucial to assess the impact of a weak static disorder on
          signal-idler correlations: disorder is in fact known to be
          responsible
          for the so-called
resonant Rayleigh scattering of pump photons\cite{Rayleigh}, which
          also produce a ring-shaped pattern in momentum space, yet
          without any quantum correlation.

In this paper, we make use of the Wigner Quantum Monte Carlo method\cite{Carusotto_QMC} for
polaritons in semiconductor microcavities to
numerically tackle these key issues.
The paper is structured as
follows. In Sec.  \ref{Method}, we present the model Hamiltonian
  and quantum Monte Carlo technique used to calculate the
  observables.
Results for the ring-shaped polariton parametric luminescence with or
  without a static disorder are reported in Sec. \ref{luminescence}.
    Corresponding numerical results for the quantum correlations
  are presented in Sec.\ref{Correlations} and then compared to a
  simplified
  quantum Langevin analytical model in Sec. \ref{io}. Finally,
  conclusions are drawn in Sec.\ref{Conclusions}.

\section{Hamiltonian and quantum Monte Carlo technique}
 \label{Method}

In this paper, we consider the quantum field Hamiltonian
introduced in Ref. \onlinecite{Iac_Crist_1}:
\begin{eqnarray}\label{Hamiltonian}
H &=&
 \int d\mathbf{x}  \sum_{ij = \{X,C \}} \hat{\Psi}^{\dag}_i(\mathbf{x})
\left[\mathbf{h}^0_{ij} +
V_i(\mathbf{x})\delta_{ij}\right]\hat{\Psi}^{\dag}_j(\mathbf{x})
\nonumber\\
&+&
 \frac{\hbar g}{2} \int d\mathbf{x}
\hat{\Psi}^{\dag}_X(\mathbf{x})  \hat{\Psi}^{\dag}_X(\mathbf{x})
\hat{\Psi}_X(\mathbf{x})\hat{\Psi}_X(\mathbf{x})\nonumber\\&&
+\int d\mathbf{x} \hbar F_{\mathbf{p}}(\mathbf{x},t)
  \hat{\Psi}_C^{\dag} (\mathbf{x}) + h.c.~,
\end{eqnarray}
where $\mathbf{x}$ is the in-plane spatial position. The field
operators $\hat{\Psi}_{X,C}(\mathbf{x})$
 respectively describe excitons and cavity photons. We assume an exciton density far
below the saturation density $n_{sat}$ \cite{Cristiano_semi}, so
the field operators obey the Bose commutation rules :
$[\hat{\Psi}_i (\mathbf{x}),\hat{\Psi}^{\dag}_j (\mathbf{x'})] =
\delta_{\mathbf{x},\mathbf{x'}} \delta_{i,j}$. The linear
Hamiltonian $\mathbf{h}^0_{ij}$ is:
\begin{eqnarray}
\mathbf{h}^0 = \hbar \left( \begin{array}{cc}{}
 \omega_X(-i \nabla)  & \Omega_R \\
\Omega_R &  \omega_C(-i \nabla) \end{array} \right)~,
\end{eqnarray}
where $\omega_C(\mathbf{k}) = \omega_C^0\sqrt{1 +
\mathbf{k}^2/k_z^2}$ is the cavity dispersion as a function of the
in-plane wavevector $\mathbf{k}$ and $k_z$ is the quantized photon
wavevector in the growth direction. The exciton dispersion is
assumed to be momentum-independent, i.e.,  $\omega_X(\mathbf{k}) = \omega_X^0$. The quantity
$\Omega_R$ is the vacuum Rabi frequency of the exciton-cavity
photon coupling. The eigenmodes of the linear Hamiltonian $h^0$
are called Lower and Upper Polaritons $(LP,UP)$. Their energies
are respectively $\hbar\omega_{LP}(k)$ and $\hbar\omega_{UP}(k)$.
 The nonlinear interaction term $g$ is due to the exciton-exciton
 collisional interactions, which are modelled by a contact
 potential\cite{Cristiano_semi}. For the sake of simplicity, we
 restrict ourselves  to the case of a circularly polarized  pump
 beam, which  allows us to ignore the spin degrees of freedom and
 the complex spin dynamics~\cite{Lagoudakis,Kavokin_spin,
Kavokin_spin_Hall}. The potential due to the static disorder is
 included in $V_{X,C}(\mathbf{x})$.


The polariton dynamics is studied by means of numerical
simulations based on the so-called Wigner quantum Monte Carlo
method, explained in detail in Ref.\onlinecite{Carusotto_QMC}.
Within this framework, the time-evolution of the quantum fields is
described by stochastic equations for the $\mathbb{C}$-number
fields $\psi_{X,C}(\mathbf{x})$. The evolution equation includes a
non-linear term due to interactions, as well as dissipation and
noise terms due to the coupling to the loss channels.  Actual
calculations are performed on a finite two-dimensional spatial
grid of $n_x \times n_y$ points regularly spaced over the
integration box of size $L_x \times L_y$. The different Monte
Carlo configurations are obtained as statistically independent
realizations of the noise terms.

Expectation values for the observables are then obtained by taking
the configuration average of the stochastic fields. As usual in
Wigner approaches~\cite{Carusotto_QMC}, the stochastic average
over noise provide expectation values for the totally symmetrized
operators, namely:
\begin{equation}
 \langle O_1 ...O_N\rangle_W \equiv \frac{1}{N!} \sum_{P} \langle \hat{O}_{P(1)} ...\hat{O}_{P(N)}\rangle~,
\end{equation}
the sum being made over all the permutations $P$ of an ensemble of
$N$ objects. Each operator $\hat{O}_a$ represents here some
quantum field component, while ${O}_a$ is the corresponding
$\mathbb{C}$-number stochastic field.\\

 The relation between real-
and momentum-space operators is:
\begin{eqnarray}
\hat{\Psi}_C (\mathbf{x}) = \frac{1}{\sqrt{L_x L_y}}
\sum_{\mathbf{k}}
e^{i\mathbf{k}\mathbf{x}}\,\hat{a}_{\mathbf{k}}~,\\
 \hat{\Psi}_X (\mathbf{x}) = \frac{1}{\sqrt{L_x L_y}} \sum_{\mathbf{k}}
e^{i\mathbf{k}\mathbf{x}}\,\hat{b}_{\mathbf{k}}~,
\end{eqnarray}
where $\hat{a}_{\mathbf{k}}$ ($\hat{b}_{\mathbf{k}}$) represents the
photonic (excitonic) destruction operator for the $\mathbf{k}$-mode,
and satisfy the usual Bose commutation rules
$[\hat{a}_{\mathbf{k}},\hat{a}^{\dag}_{\mathbf{k}'}] =
[\hat{b}_{\mathbf{k}},\hat{b}^{\dag}_{\mathbf{k}'}] =
\delta_{\mathbf{k},\mathbf{k}'}$.
The expectation value of the
in-cavity photon population $\hat{n}_{\mathbf{k}}=
\hat{a}^{\dag}_{\mathbf{k}}\,\hat{a}_{\mathbf{k}}$  in the
$\mathbf{k}$-mode reads:
\begin{equation}
\langle \hat{n}_{\mathbf{k}} \rangle  = \frac{1}{2}\,\langle
\hat{a}^{\dag}_{\mathbf{k}}\hat{a}_{\mathbf{k}} +
\hat{a}_{\mathbf{k}}\hat{a}^{\dag}_{\mathbf{k}} \rangle -
\frac{1}{2}=
\overline{|\alpha_{\mathbf{k}}|^2} - \frac{1}{2} ~,
\end{equation}
where the overlined quantities are stochastic configuration averages,
and
$\alpha_{\mathbf{k}}$
is the
 $\mathbb{C}$-number stochastic field value corresponding to the
 operator
$\hat{a}_{\mathbf{k}}$
Because of the weak, but still finite transmittivity of the cavity
mirrors, all observables for the in-cavity field
transfer~\cite{Carusotto_QMC,Walls} into the corresponding ones for
the extra-cavity luminescence at the same in-plane momentum $\mathbf{k}$.

\section{Results for the ring-shaped luminescence}
\label{luminescence}
\subsection{In the absence of disorder}
\label{Disorderless}

\begin{figure}[!ht]
\begin{center}
\includegraphics[width=8cm]{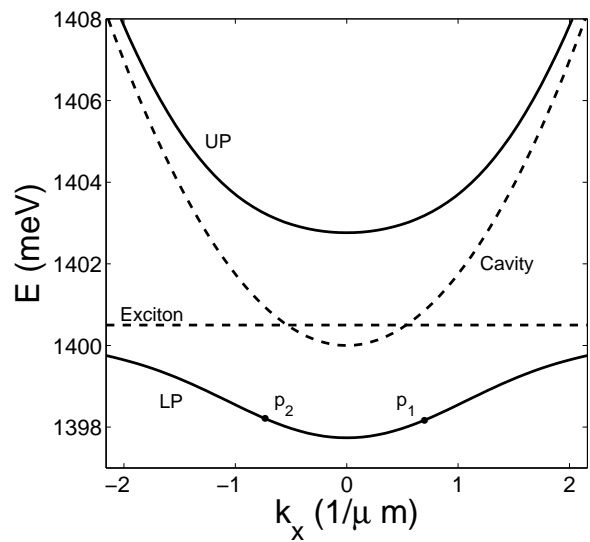}
\caption{\label{dispersion}Solid lines: energy dispersion of the
Lower and Upper Polariton branches. Dashed lines: the bare quantum
well exciton and cavity photon dispersions. The filled circles
indicate the wavevectors of the two pumps ($p_1$ and $p_2$). Note
that in the chosen configuration, the photonic fraction of the
 Lower Polariton at the pump wavevector is $\simeq$ 0.46.
 Cavity parameters: $\hbar\Omega_R = 2.5$ meV, $\hbar\omega_C^0
=$ 1400 meV, $\hbar \omega_X^0 = 1400.5$
 meV, $k_z = 20\,\mu\textrm{m}^{-1}$. Pump parameters: $k_p
 =0.6981\,\mu\textrm{m}^{-1}$, $\omega_p=\omega_{LP}(\mathbf{k}_p)=1398.2$ meV. }
\end{center}
\end{figure}

In this work, we will consider the following excitation field:
\begin{eqnarray}
 F_{\mathbf{p}}(\mathbf{x},t) = F_p \left(e^{-i{k}_p {x}} + e^{i{k}_p {x}}\right)e^{-i \omega_p t}~.
\end{eqnarray}
This field describes two identical monochromatic plane-wave pumps
with opposite wavevectors oriented along the x-axis. Both beams
have the same values for the amplitude $F_p$ and the frequency
$\omega_p$. This latter is chosen to be  resonant with the
LP-branch, i.e. $\omega_p = \omega_{LP}(\mathbf{k}_p)$. Fig.
\ref{dispersion} depicts the dispersion of the polariton branches
and the position of the pump wavevectors. The scattering process
between a pair of $\pm\mathbf{k}_p$ pump polaritons via the
non-linear interactions, gives  rise to a pair of signal/idler
polaritons of opposite wavevectors $\pm\mathbf{k}$. Modulo the
weak blue-shift of the modes due to interactions, the
energy-momentum conservation (phase-matching) is trivially
fulfilled if $\omega_s = \omega_i = \omega_p$ and $|\mathbf{k}| =
{k}_p$, that is on the $|\mathbf{k}| = {k}_p$ parametric luminescence ring.

\begin{figure}[!ht]
\begin{center}
\includegraphics[width=8.5cm]{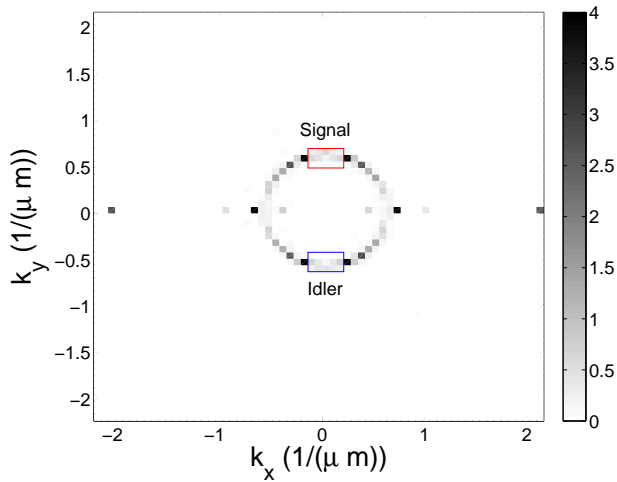}
\caption{\label{cavity_photon_kx_ky_1} QMC results for the
in-cavity photon population $n_{\mathbf{k}}$. Pump
  amplitude  $F_p /\gamma= 5\,\mume^{-1}$ (just  below the parametric instability threshold).
Number of Monte Carlo configurations : 330.
 The two rectangles denote the areas where the signal and idler are
 integrated.
Cavity parameters: $\hbar\Omega_R = 2.5$ meV, $\hbar\omega_C^0 =$
1400 meV, $\hbar \omega_X = 1400.5$ meV, $k_z = $20 $\mu$m$^{-1}$,
$\hbar\Omega_R = 2.5$~meV,  $\hbar\gamma_{C,X} =
 \hbar\gamma = 0.1$ meV, $\hbar\,g=$ 1.10$^{-2}$ meV.$\mu$m$^{-2}$. Pump parameters: $k_p=$ 0.6981
 $\mu$m$^{-1}$, $\omega_p=\omega_{LP}(\mathbf{k_p})= 1398.2$ meV.
Integration box size $L_x = L_y = 90$ $\mu$m with $n_x = n_y =
 64$ points; integration time step $dt=1.7$ fs. Using such
 a short time step has been necessary in order to obtain sufficient
 numerical precision on  fourth-order field correlation functions.}
\end{center}
\end{figure}

Fig. \ref{cavity_photon_kx_ky_1} shows the numerical results for
the stationary state photon population inside the cavity for a
value of the pump power below the parametric oscillation
threshold:  the ring-shaped parametric luminescence pattern is
apparent. The interaction-induced blue shift of the polariton
modes is responsible for the ring radius being slightly smaller
than $k_p$. Other interesting features can be observed in addition
to the main ring: the strong spots at $\mathbf{k} = \pm
3\mathbf{k}_p$ are due to four-wave mixing processes
$(\mathbf{k}_p,\mathbf{k}_p) \rightarrow (\pm
3\mathbf{k}_p,\mp\mathbf{k}_p)$; because of the stimulated nature
of the underlying process, these spots fully inherit the coherence
of the pump beams. Some luminescence is also observed along the
x-axis in the vicinity of $\mathbf{k}_p$. Parametric scattering
processes involving polaritons from the same pump beam
$(\mathbf{k}_p,\mathbf{k}_p) \rightarrow  (\mathbf{k}_p +
\delta\mathbf{k} ,\mathbf{k}_p - \delta \mathbf{k})$ with
$|\delta \mathbf{k}| \ll |\mathbf{k}_p|$ are
responsible for this emission. As the pump beams are not tuned at
the so-called magic angle, this emission is much weaker than the one
on the ring.

In the following, we will focus our attention on signal-idler
pairs with wavevectors on the ring, and close to the y-axis ($k_x
\simeq 0$). To minimize discretization effects, we will average
the signal/idler observables on the rectangular areas
$\mathcal{D}_{s,i}$ sketched in Fig. \ref{cavity_photon_kx_ky_1},
which indeed contain quite a number of pixels. The corresponding
photon population operators $\hat{N}_{s,i}$ are defined as:
\begin{eqnarray}
 \hat{N}_{s,i}  = \sum_{\mathbf{k}\in \mathcal{D}_{s,i}} \hat{a}^{\dag}_{\mathbf{k}}\hat{a}_{\mathbf{k}} = N_D\hat{n}_{s,i}~,
\end{eqnarray}
where $N_{\mathcal{D}}$ is the number of modes inside
$\mathcal{D}_{s,i}$ and $\hat{n}_{s,i}$ are the average photon
population operators. In term of the stochastic field, these
latter read:
\begin{equation}
 \langle \hat{n}_{s,i} \rangle = \frac{1}{N_D}
\sum_{\mathbf{k}\in \mathcal{D}_{s,i}}
\left(\overline{|\alpha_{\mathbf{k}}|^2}
- \frac{1}{2}\right)~.\\
\end{equation}

\begin{figure}[!ht]
\begin{center}
\includegraphics[width=9cm]{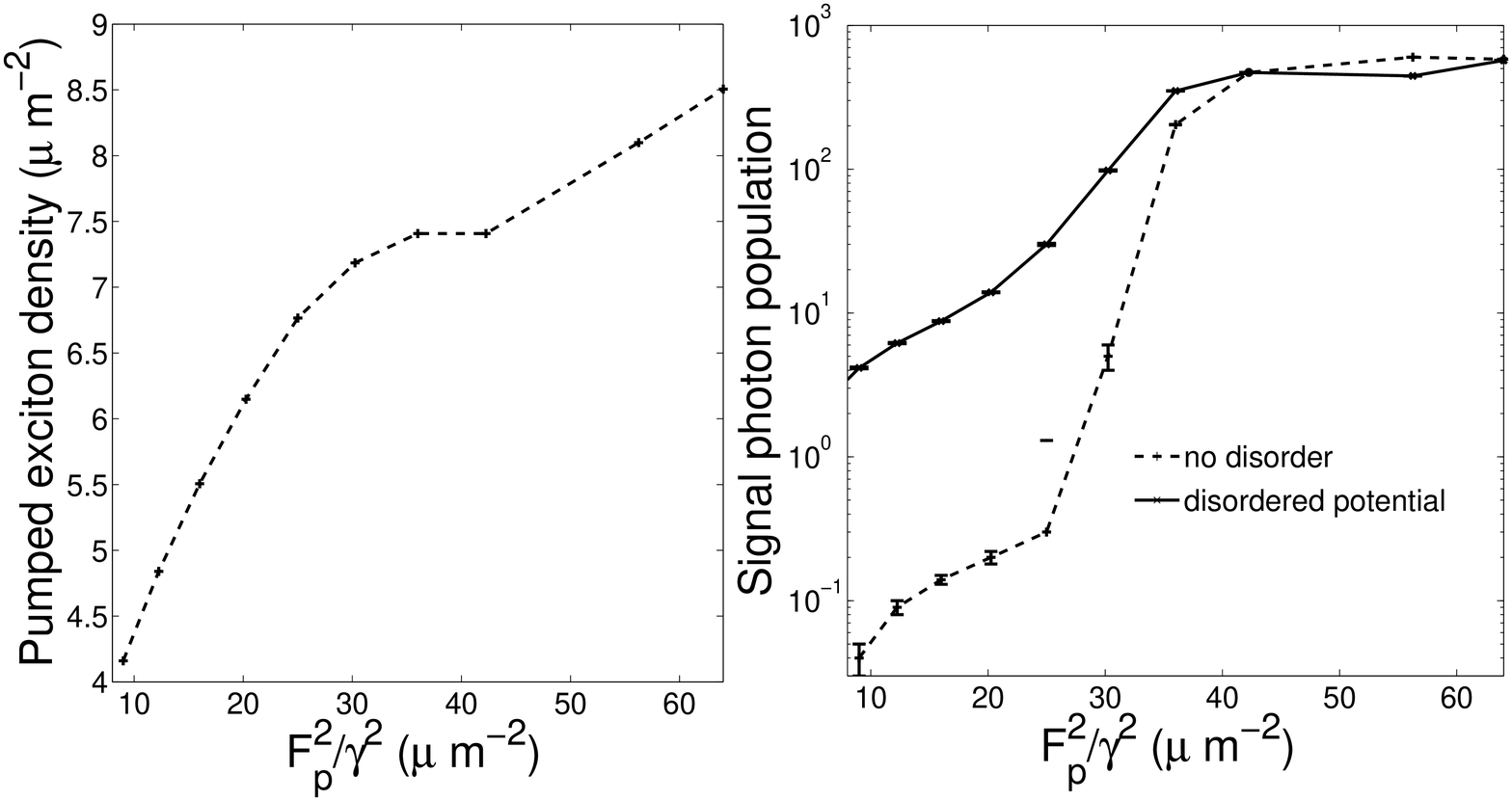}
\caption{\label{pop_QMC} Left panel: QMC
results for the density of the pump excitons $\rho_{p}
= \langle \hat{b}_{\mathbf{k}_p}^{\dag}\hat{b}_{\mathbf{k}_p}\rangle / (L_x
L_y )$. Right panel: signal/idler photon populations $n_{s,i}$ in the
presence (solid line) or absence (dashed line) of the disordered potential. Same cavity and integration parameters as in
Fig.\ref{cavity_photon_kx_ky_1}. }
\end{center}
\end{figure}

The density of pump excitons $\rho_{p}= \langle
\hat{b}_{\pm\mathbf{k}_p}^{\dag}\hat{b}_{\pm\mathbf{k}_p}\rangle /
(L_x L_y )$ and the signal/idler populations $n_{s,i}$ are shown
as a function of pump power in the left and right panels of Fig.
\ref{pop_QMC} respectively. As previously
discussed\cite{Carusotto_QMC,threshold}, the pump density $\rho_p$
smoothly increases up to the threshold for parametric oscillation.
The sub-linear dependence on power stems from the optical limiting
effect due to the blue-shift of the $\pm\mathbf{k}_p$ modes by the
repulsive interactions\cite{Iac_Crist_1}. Around the threshold at
$F_p/\gamma\approx 5.75\,\mume^{-1}$, $\rho_p$ shows a downward
kink, while the signal/idler populations have a sudden increase.
For the realistic parameters used here, note how the density of
excitons at the instability threshold remains moderate, and much
lower than the exciton saturation density, $\rho_{\mathbf{k}_p} <
10^9 \,{\rm   cm}^{-2}\ll \rho_{sat}$. This shows the efficiency
of the considered parametric process.

\begin{figure}[!ht]
\begin{center}
\includegraphics[width=8.5cm]{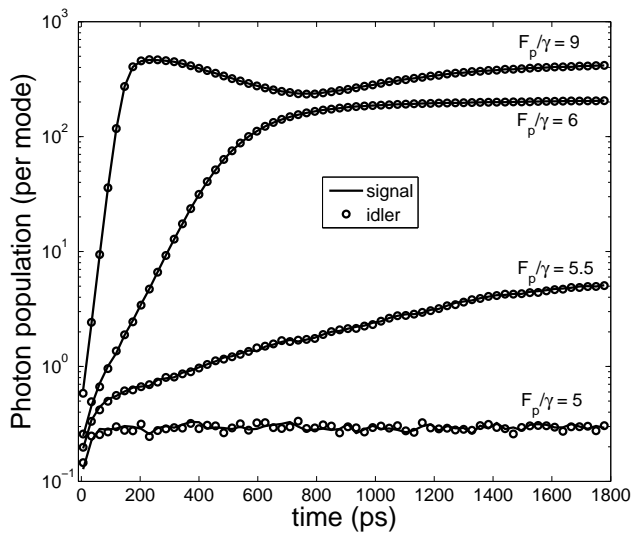}
\caption{\label{pop_time} QMC results for $\langle \hat{n}_s\rangle$ (Solid line) and
$\langle \hat{n}_i\rangle$ (circles) as a function of time (ps) for
 various pump amplitudes across the parametric oscillation threshold:
  $F_p /\gamma = 5\,\mume^{-1}$  (330 configurations),
 $F_p /\gamma= 5.5\,\mume^{-1}$  (340 configurations), $F_p / \gamma= 6\,\mume^{-1}$
 (180 configurations) and  $F_p / \gamma = 9\,\mume^{-1}$  (60 configurations).
Same cavity and integration parameters as in
Fig.\ref{cavity_photon_kx_ky_1}. }
\end{center}
\end{figure}

\begin{figure}[!ht]
\begin{center}
\includegraphics[width=8.5cm]{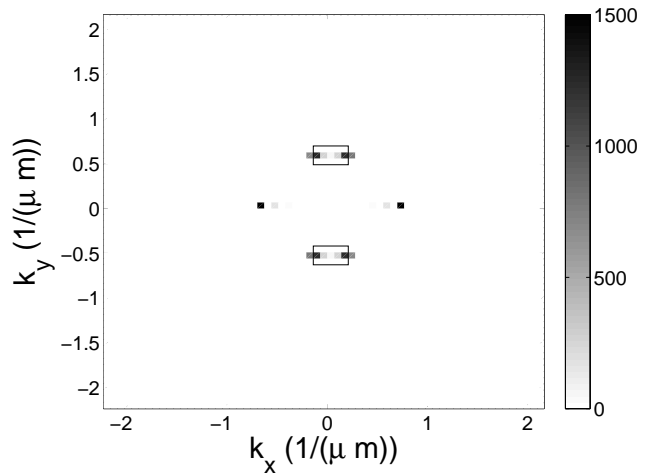}
\caption
{\label{cavity_photon_kx_ky_3} QMC results for the
  in-cavity photon population $n_{\mathbf{k}}$
  for a pump
amplitude above the parametric oscillation
  threshold $F_p /\gamma = 6\,\mume^{-1}$. Number of Monte Carlo configurations
: 180. Same cavity and integration parameters as in
Fig.\ref{cavity_photon_kx_ky_1}.}
\end{center}
\end{figure}

In Fig. \ref{pop_time}, we can see the temporal spontaneous build-up of the signal
and idler luminescence starting from the vacuum fluctuations. For
$F_p/\gamma = 5\,\mume^{-1}$, the population of the signal/idler modes is
still small $n_{s,i} \ll 1$, while stimulated parametric
scattering starts to be effective for $F_p/\gamma = 5.5\,\mume^{-1}$ when the
occupation number is comparable or larger than $1$. The parametric
oscillation threshold has already been crossed for $F_p/\gamma =
6\,\mume^{-1}$. While the emission ring below threshold has the almost
homogeneous intensity profile shown in Fig.
\ref{cavity_photon_kx_ky_1}, a symmetry breaking takes place above
the threshold: a few modes are selected by mode competition
effects, and a macroscopic population concentrates into them, as
shown in Fig.\ref{cavity_photon_kx_ky_3}.
It is interesting to note that that the number of Monte Carlo
configurations needed for obtaining a given precision in the
configuration average strongly depends on the regime under
examination: as expected, much less simulations are required above
the threshold.

\subsection{In the presence of static disorder}
\begin{figure}[!ht]
\begin{center}
\includegraphics[width=8cm]{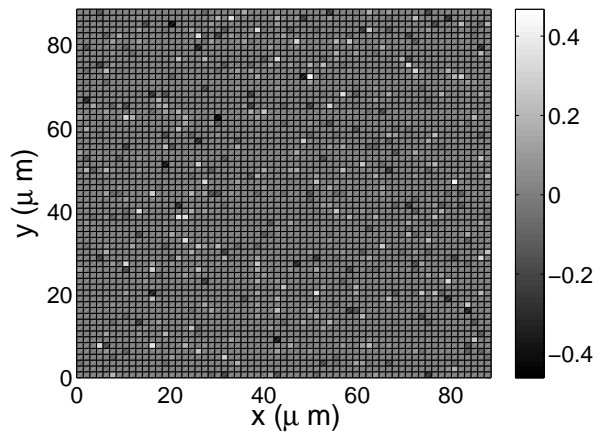}
\caption{\label{potential_sample}Disordered photonic potential (in
meV/$\hbar$) used for the simulations of Fig.
 \ref{pop_QMC},\ref{cavity_photon_kx_ky_4}, and
\ref{sum_difference_disordered}. }
\end{center}
\end{figure}

\begin{figure}[!ht]
\begin{center}
\includegraphics[width=8.5cm]{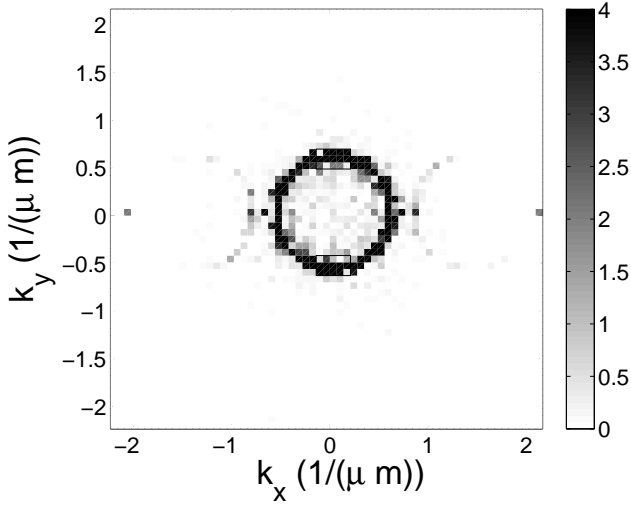}
\caption{\label{cavity_photon_kx_ky_4} QMC result for the
in-cavity photon population $n_{\mathbf{k}}$ in the
  presence of the disordered potential shown in
  Fig. \ref{potential_sample}. The
image saturates the gray scale. Pump amplitude $F_p / \gamma= 5\,\mume^{-1}$.
 Number of Monte Carlo configurations : 50. Same cavity and
integration parameters as in Fig.\ref{cavity_photon_kx_ky_1}.}
\end{center}
\end{figure}

The results in Fig. \ref{cavity_photon_kx_ky_1}, \ref{pop_time}
and \ref{cavity_photon_kx_ky_3} have been obtained in the absence
of static disorder, i.e. for $V_C = V_X=0 $. An arbitrary
potential can be easily included in our calculations. As a
specific example, we have considered the disordered photonic
potential reported in Fig. \ref{potential_sample}, consisting of a
random ensemble of photonic point defects~\cite{note} . The
corresponding emission pattern is shown in Fig.
\ref{cavity_photon_kx_ky_4} for the same pump parameters as in the
clean system of Fig. \ref{cavity_photon_kx_ky_1}. The main effect
of the disorder appears to be a significantly enhanced intensity
on the luminescence ring. This occurs because of the resonant
Rayleigh scattering of each of the pumps. Note also the weak
``eight''-shaped pattern\cite{Cristiano_semi} due to the
parametric amplification of the resonant Rayleigh scattering ring.

The signal population as a function of the pump power is plotted
in the right panel of Fig. \ref{pop_QMC}:  below threshold, the photon
population in the presence of disorder is much larger than in the
clean system. On the other hand, the difference between the two
populations is much less important above threshold when the
non-linear stimulated parametric scattering dominates over the
linear Rayleigh scattering processes. Despite the very different
low intensity behaviour, the threshold is reached in both cases at
values close to $F_p/\gamma \simeq 5.5\,\mume^{-1}$.

\section{Quantum Correlations}
\label{Correlations}

 In the present section we study the correlation properties of
the signal and idler emissions. For the sake of simplicity, we
restrict our attention here to those fluctuations which are
associated to the intrinsic losses of the parametrically emitting
system, and we neglect all other possible noise sources that may
appear in actual experimental setups, e.g. pump intensity
fluctuations. To characterize the quantum nature of the
correlations\cite{Mandel} between the signal and idler modes, it
is useful to consider the quantity $\hat{N}_{\pm} = \hat{N}_s \pm
\hat{N}_i$ i.e. the sum and difference of the signal and idler
photon population. The corresponding normalized noise
$\sigma^{\pm}$ reads: \onecolumngrid
\begin{eqnarray}
{\sigma^{\pm}} = \frac{ \langle  \hat{N}_{\pm}^2 \rangle
 -\langle \hat{N}_{\pm} \rangle^2}
{\langle \hat{N}_+ \rangle} = \frac{ \langle  \hat{N}_{s}^2
\rangle -\langle  \hat{N}_{s} \rangle^2 + \langle  \hat{N}_{i}^2
\rangle -\langle  \hat{N}_{i} \rangle^2 \pm  2 \left(\langle
\hat{N}_{i} \hat{N}_{s} \rangle - \langle \hat{N}_{s}\rangle
\langle\hat{N}_{i} \rangle\right)
 }
{\langle   \hat{N}_{i} \rangle + \langle  \hat{N}_{s} \rangle}~.
\end{eqnarray}
Hence, the fourth-order moments of the fields $\langle \hat{N}_j \hat{N}_k \rangle
= \sum_{\mathbf{k}\in \mathcal{D}_j, \mathbf{k'}\in \mathcal{D}_k}
 \langle \hat{n}_{\mathbf{k}}\hat{n}_{\mathbf{k'}}
\rangle$ (where $j,k\in\{s,i\}$)  play a key-role in the
determination of the quantum behaviour of the system.
  In terms of the averaged stochastic quantities we have:
\begin{eqnarray}
\langle \hat{N}_{j}^2 \rangle &=& \frac{1}{6}\sum_{\mathbf{k}\in
\mathcal{D}_{j}} \left\langle
\hat{a}_{\mathbf{k}}^{\dag}\hat{a}_{\mathbf{k}}^{\dag}
\hat{a}_{\mathbf{k}} \hat{a}_{\mathbf{k}}
+\hat{a}_{\mathbf{k}}^{\dag}\hat{a}_{\mathbf{k}}
\hat{a}_{\mathbf{k}}^{\dag} \hat{a}_{\mathbf{k}}
+\hat{a}_{\mathbf{k}}^{\dag}\hat{a}_{\mathbf{k}}
\hat{a}_{\mathbf{k}} \hat{a}_{\mathbf{k}}^{\dag}
+\hat{a}_{\mathbf{k}}\hat{a}_{\mathbf{k}}^{\dag}
\hat{a}_{\mathbf{k}}^{\dag} \hat{a}_{\mathbf{k}}
+\hat{a}_{\mathbf{k}}\hat{a}_{\mathbf{k}}^{\dag}
\hat{a}_{\mathbf{k}} \hat{a}_{\mathbf{k}}^{\dag}
+\hat{a}_{\mathbf{k}}\hat{a}_{\mathbf{k}}
\hat{a}_{\mathbf{k}}^{\dag} \hat{a}_{\mathbf{k}}^{\dag}
\right\rangle  -\frac{1}{2}\langle
\hat{a}_{\mathbf{k}}^{\dag}\hat{a}_{\mathbf{k}}
+\hat{a}_{\mathbf{k}} \hat{a}_{\mathbf{k}}^{\dag} \rangle\nonumber\\
&& \phantom{\frac{1}{6}\sum_{\mathbf{k}\in \mathcal{D}_{j}}}
+\sum_{ \mathbf{k},\mathbf{k'}\in \mathcal{D}_j}
 \langle\hat{n}_{\mathbf{k}}\rangle \langle\hat{n}_{\mathbf{k'}}\rangle
- \sum_{ \mathbf{k}\in \mathcal{D}_j}
 \langle\hat{n}_{\mathbf{k}}\rangle \langle\hat{n}_{\mathbf{k}}\rangle
\nonumber\\
&=&
\sum_{\mathbf{k}\in
\mathcal{D}_j}\left(
\overline{|\alpha_{\mathbf{k}}|^4}
 -
\overline{|\alpha_{\mathbf{k}}|^2}
\right)+ \left(\sum_{\mathbf{k}\in \mathcal{D}_j }\left(
\overline{|\alpha_{\mathbf{k}}|^2}
-\frac{1}{2} \right)
\right)^2 - \sum_{\mathbf{k}\in \mathcal{D}_j }\left(
\overline{|\alpha_{\mathbf{k}}|^2}
-\frac{1}{2}\right)^2~,
\end{eqnarray}
\twocolumngrid
 with $j = (i,s)$.
The intensity correlation between signal and idler modes is:
\onecolumngrid
\begin{eqnarray}
 \langle \hat{N}_{s} \hat{N}_{i} \rangle &=&
\frac{1}{4} \sum_{\stackrel{ \scriptstyle
\mathbf{k}\in\mathcal{D}_s} { \scriptstyle
\mathbf{k}'\in\mathcal{D}_i}} \left\langle (
\hat{a}_{\mathbf{k}}^{\dag} \hat{a}_{\mathbf{k}} +
\hat{a}_{\mathbf{k}} \hat{a}_{\mathbf{k}}^{\dag} - 1 ) \right.
\left. ( \hat{a}_{\mathbf{k'}}^{\dag} \hat{a}_{\mathbf{k'}} +
\hat{a}_{\mathbf{k'}} \hat{a}_{\mathbf{k'}}^{\dag} - 1
)\right\rangle
 \\
  &=&
\sum_{\mathbf{k}\in \mathcal{D}_{s}}\sum_{\mathbf{k}'\in \mathcal{D}_{i}}
\overline{|\alpha_{\mathbf{k}}|^2\,|\alpha_{\mathbf{k}'}|^2}
- \frac{N_D}{2}\left(
\sum_{\mathbf{k}\in \mathcal{D}_{s}}
\overline{| \alpha_{\mathbf{k}}|^2}
+ \sum_{\mathbf{k}'\in \mathcal{D}_{i}}
\overline{| \alpha_{\mathbf{k}'}|^2}
\nonumber
-\frac{N_D}{2} \right)~.
\end{eqnarray}
\twocolumngrid
 For uncorrelated and shot-noise limited signal and idler beams,
one would have $\sigma^{\pm}= 1$: this value is the so-called
Standard Noise Limit\cite{Walls}. Having $\sigma^- < 1$ means that
non-classical correlations exist between signal and idler, in
particular a squeezing of the difference intensity
noise\cite{Fabre_Treps,Fabre_Houches}.  As the polariton states
are half-photon half-exciton, the optimal noise reduction of the
photon field is reduced by half with respect to an ideal
$\chi^{(2)}$ purely photonic parametric oscillator
system\cite{Laurat03}; noise reduction does not concern in fact
the photon field taken independently, but rather the whole
polariton field. As long as we neglect multiple scattering and
disorder effects, this is the main difference compared to standard
$\chi^{(2)}$ parametric oscillators; an analytical model for these
issues will be provided in Sec.\ref{io}. Note that throughout all
the present paper we are interested in one-time correlations: the
difference noise is therefore integrated over all the frequencies
and no frequency filtering is considered. Some frequency filtering
around $\omega = \omega_p$ would purify the squeezing as the
quantum correlations are largest around this value of the
frequency\cite{Fabre_Cargese}.

\begin{figure}[!ht]
\begin{center}
\includegraphics[width=9cm]{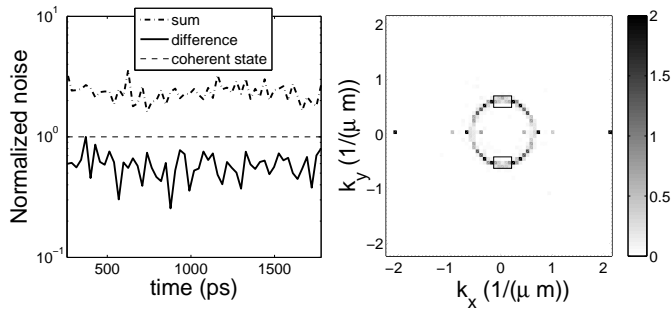}
\caption{\label{sum_difference_ex}Left panel: QMC results for the
  time-evolution of the normalized photonic sum and difference noises
  $\sigma^{\pm}$   in the absence of disorder.
Right panel: in-cavity photon population $n_{\mathbf{k}}$ at $t
  = 1800$~ps. Pump amplitude  $F_p /\gamma= 5\,\mume^{-1}$. Number of Monte Carlo
  configurations: 330. Same cavity and integration parameters as in
Fig.\ref{cavity_photon_kx_ky_1}.}
\end{center}
\end{figure}

\begin{figure}[!ht]
\begin{center}
\includegraphics[width=9cm]{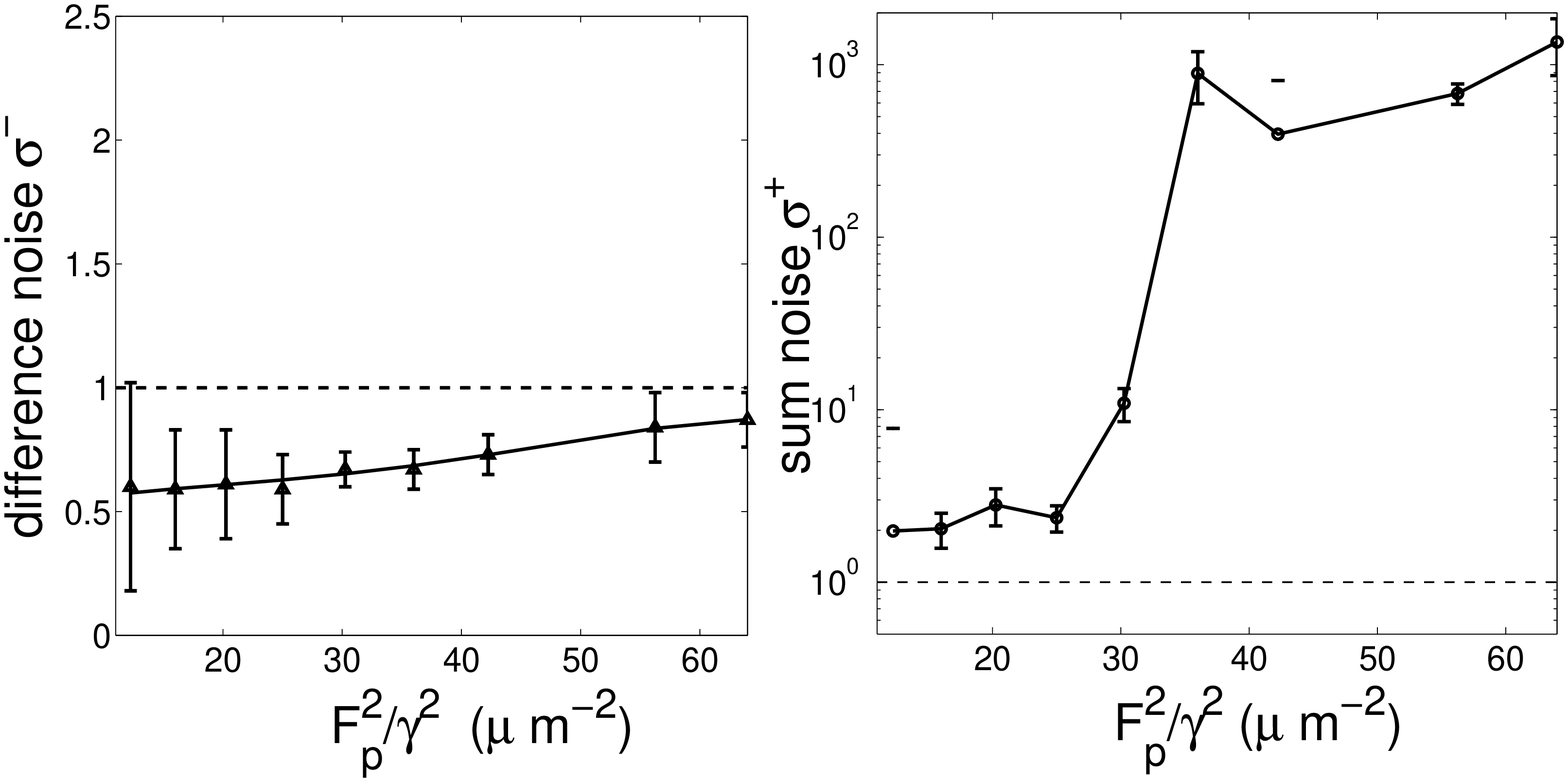}
\caption{\label{sum_difference}QMC results for the normalized photonic
 sum and difference noises $\sigma^\pm$ as a function of the pump power in the absence of
 disorder.
  The difference noise is fitted by a polynomial curve, while the
 line in the plot for $\sigma^+$ is a guide for the eye. Same cavity
 and integration parameters as in
Fig.\ref{cavity_photon_kx_ky_1}.}
\end{center}
\end{figure}

In Fig. \ref{sum_difference_ex} we have plotted the time
dependance of the normalized noises $\sigma^{\pm}$ for a given
Monte Carlo realization and in the absence of disorder. The
stationary-state average values for the same quantities are
plotted in Fig. \ref{sum_difference} as a function of the pump
intensity.
Quantum correlations $\sigma^-<1$ exist in the difference noise at low
intensities, while it monotonically increases towards $\sigma^-=1$ for
higher intensities, making the signal/idler correlations almost purely
classical well above threshold\cite{Fabre_Treps}. No specific feature
is found in this quantity at the threshold.

On the other hand, the sum noise $\sigma^+$ is always above the
standard noise limit, and shows a sudden increase at the parametric
threshold.
The fact that well above the threshold it does not go back to the
standard noise is due to the presence of several competing parametric
oscillation modes. Depending on whether the oscillating modes lay
inside or outside the selected regions ${\mathcal D}_{s,i}$, the
signal/idler populations $n_{s,i}$ vary between $0$ and their maximum
value, while remaining almost equal to each other. This implies that
the sum noise $\sigma^+$ is large, of the order of the signal/idler
populations $n_{s,i}$ while the difference noise $\sigma^-$ remains
small.

\begin{figure}[!ht]
\begin{center}
\includegraphics[width=9cm]{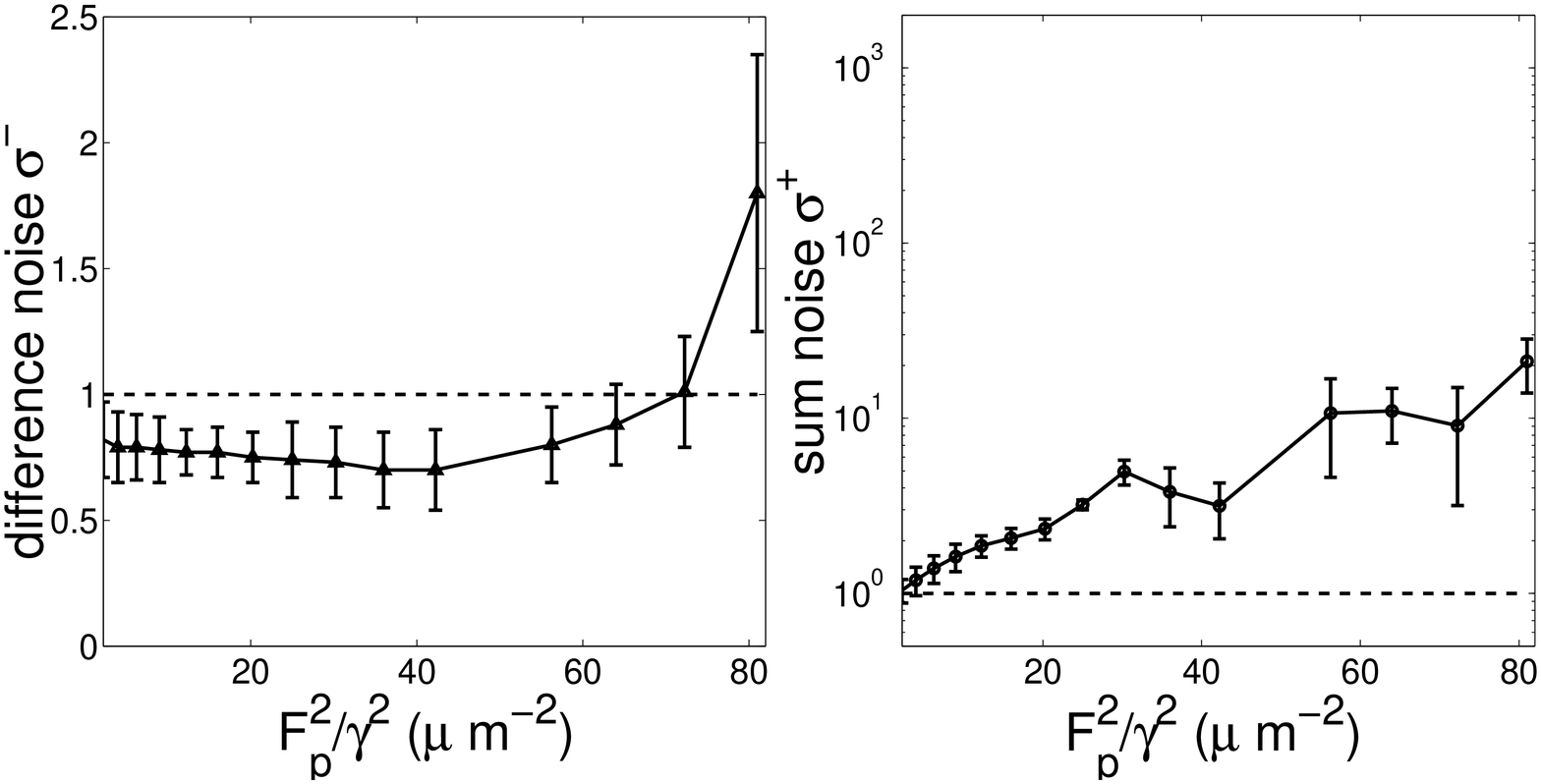}
\caption{\label{sum_difference_disordered}QMC results for  $\sigma^\mp$
  as a function of the pump intensity in the presence of disorder.
The lines are a guide for the eye.
 Same cavity and integration parameters as in
Fig.\ref{cavity_photon_kx_ky_1}. }
\end{center}
\end{figure}

In Fig.\ref{sum_difference_disordered}, we have analyzed the sum
and difference noise in the presence of a disordered potential.
For the same value of pump intensity, the difference noise
$\sigma^{-}$ is now somehow larger than in the absence of
disorder: the resonant Rayleigh scattering creates in fact
unpaired photons into the luminescence ring and deteriorates the
pair correlations between the signal and the idler. For very low
intensities, the dominant contribution comes from the Rayleigh
scattering processes implying that both the sum and the difference
noises have to tend toward the Standard Noise Limit
$\sigma^\pm=1$. Because of the competition between the Rayleigh
and the parametric scattering, the difference noise $\sigma^-$
attains its minimum in the vicinity of the threshold and then
increases because of the increasing noise of the two beams. As
disorder is able to mix the modes respectively inside and outside
the selected regions ${\mathcal D}_{s,i}$, the difference noise
can grow above $\sigma^-=1$ at high pump powers. For the same
reason, the sum noise $\sigma^+$ has a weaker growth above the
threshold than in the absence of disorder. This physical
interpretation of the role of the disorder has been confirmed by
several other simulations (not shown) performed with different
realizations of the disordered potential.
\\

\section{Simplified analytical model}
\label{io}

The aim of this section is to compare the results of the complete
 Quantum Monte-Carlo calculations to a
 simplified input-output analytical model based on a linearization of
 the Hamiltonian. This is done by  treating the
 intense pump as a classical, undepleted, field, i.e. replacing the
 pump mode operators with their mean-field expectation values.
Obviously, this approximation is valid only well below the parametric
 oscillation threshold.
 Concentrating our attention on those processes which satisfy the
phase-matching condition and neglecting all the non-resonant
others,
 we can write the linearized Hamiltonian in the following, simplified form:
\begin{multline}
\mathcal{H}  =
 \sum_{\mathbf{k}\neq \mathbf{k}_p} \left[ \hbar\omega_C(\mathbf{k}) \hat{a}_{\mathbf{k}}^{\dag} \hat{a}_{\mathbf{k}}
  + \hbar{\tilde \omega_X} \hat{b}_{\mathbf{k}}^{\dag} \hat{b}_{\mathbf{k}}
 \right. \\
+ \hbar\Omega_R \left(\hat{b}_{\mathbf{k}}^{\dag} \hat{a}_{\mathbf{k}} +
 \hat{a}_{\mathbf{k}}^{\dag} \hat{b}_{\mathbf{k}}\right) \\
\left. + \hbar \left(
\hat{b}_{\mathbf{k}}^{\dag}\hat{b}_{-\mathbf{k}}^{\dag}\kappa +
\hat{b}_{\mathbf{k}}\hat{b}_{-\mathbf{k}}\kappa^*\right)\right]~,
\end{multline}
where $\hat{b}_{\mathbf{k}}$ is the exciton creation operator,
${\tilde \omega_X} = \omega_X^0 +\frac{2g}{L_xL_y}(
|\mathcal{P}_1|^2  + |\mathcal{P}_2|^2)$ is the blue-shifted
exciton frequency because of interactions, and  $\kappa =
\frac{g}{L_xL_y} \mathcal{P}_1 \mathcal{P}_2 $ is the effective
parametric interaction constant in terms of the pump fields $
\mathcal{P}_{1,2} e^{-i\omega_p t}= \langle
\hat{b}_{\pm\mathbf{k}_p}(t) \rangle$.
 Taking the
 standard vacuum as the initial state of the photon and exciton
 fields, the expectation  values  of the quantum Langevin forces are:
\begin{eqnarray}
\langle \tilde{F}_{cav,\mathbf{k}} [\omega] \tilde{F}^{\dag}_{cav,\mathbf{k}'} [\omega]\rangle
 &=& 4\pi \Gamma_{cav}[\omega] \delta(\omega - \omega')\delta_{\mathbf{k},\mathbf{k}'}~,\ \\
\langle \tilde{F}_{exc,\mathbf{k}} [\omega] \tilde{F}^{\dag}_{exc,\mathbf{k}'} [\omega]\rangle
 &=& 4\pi \Gamma_{exc}[\omega] \delta(\omega -
 \omega')\delta_{\mathbf{k},\mathbf{k}'}~,\
\end{eqnarray}
where $\Gamma_{j}[\omega]$ is the complex broadening due to the
coupling to the external bath. Since the relevant spectral domain
 in the degenerate parametric process is concentrated around $\omega_p$, we are
allowed to simplify the treatment by taking frequency independent
linewidth $\Gamma_{ph,exc}[\omega] = \gamma_{C,X}/2$. The quantum
Langevin equations in frequency space read:
\onecolumngrid
\begin{eqnarray}
\mathcal{M}_{\mathbf{k},\omega,\omega_p}\left( \begin{array}{c}
\tilde{a}_{\mathbf{k}}[\omega]\\
\tilde{b}_{\mathbf{k}}[\omega]\\
\tilde{a}^{\dag}_{-\mathbf{k}}[2\omega_p -\omega]\\
\tilde{b}^{\dag}_{-\mathbf{k}}[2\omega_p -\omega]
\end{array}\right)
=- i \left( \begin{array}{c}
\tilde{F}_{cav,\mathbf{k}}[\omega]\\
\tilde{F}_{exc,\mathbf{k}}[\omega]\\
\tilde{F}_{cav,-\mathbf{k}}^{\dag}[2\omega_p -\omega]\\
\tilde{F}_{exc,-\mathbf{k}}^{\dag}[2\omega_p -\omega]
\end{array}\right)\begin{array}{c}\\ \\ \\ \\~,\end{array}
\end{eqnarray}
with the matrix $\mathcal{M}_{\mathbf{k},\omega,\omega_p}$ defined
for $i= X,C$ as:

\begin{eqnarray}
\mathcal{M}_{\mathbf{k},\omega,\omega_p}=
\left( \begin{array}{cccc}
 \Delta_C (\omega) -i\gamma_C/2 & \Omega_R & 0 & 0 \\
\Omega_R & \Delta_X(\omega) -  i\gamma_X/2 &  0  & \kappa \\
0 & 0 &  -\Delta_C(\omega-2\omega_p) -i\gamma_C/2 &  -\Omega_R \\
0 & -\kappa^*& -\Omega_R & - \Delta_X(\omega-2\omega_p) -i \gamma_X/2
\end{array}\right)~,
\end{eqnarray}
\twocolumngrid

in terms of $\Delta_{i}(\omega) = \omega_i - \omega$ .

The relation between the time dependent and frequency
dependent operators is:
\begin{equation}
\hat{a}_{\mathbf{k}} (t) = \int \frac{d\omega}{2\pi}
\tilde{a}_{\mathbf{k}}[\omega] e^{i\omega t},
\end{equation}
where $\tilde{a}_{\mathbf{k}}[\omega]$ is the component at $\omega$ of
the photonic destruction operator for the $\mathbf{k}$-mode.
 In the following, we will call
$\mathcal{G}_{\mathbf{k},\omega,\omega_p} =
-i\mathcal{M}_{\mathbf{k},\omega,\omega_p}^{-1}$. The signal
photon population operator $\hat{n}_{s}(t)$ inside the cavity can
be written as:
\begin{equation}
 \hat{N}_{s}(t) = \sum_{\mathbf{k}\in \mathcal{D}_s} \iint \frac{d\omega}{2\pi}\frac{d\omega'}{2\pi}
 \tilde{a}_{\mathbf{k}}^{\dag}[\omega]\,\tilde{a}_{\mathbf{k}}[\omega']\, e^{-i(\omega -\omega')t}~,
\end{equation}
which leads to:
\begin{equation}
 \langle N_{s} \rangle =
 \sum_{\mathbf{k}\in \mathcal{D}_s}\int \frac{d\omega}{2\pi} \left( \gamma_C |\mathcal{G}_{13}|^2[\omega]
+ \gamma_X
|\mathcal{G}_{14}|^2[\omega]\right).
\end{equation}

To calculate the sum and difference noise, the second order momenta
$\langle\hat{N}_s^2\rangle-  \langle \hat{N}_s\rangle^2$, $\langle\hat{N}_i^2\rangle- \langle \hat{N}_i\rangle^2$
 and $\langle \hat{N}_i \hat{N}_s\rangle + \langle \hat{N}_s \hat{N}_i\rangle
 - 2\langle \hat{N}_i\rangle\langle \hat{N}_s\rangle$ are needed.
After some algebra, we obtain the final expressions (for $j=(s,i)$):

\onecolumngrid
\begin{eqnarray}
\langle \hat{N}_j^2\rangle -  \langle \hat{N}_j\rangle^2 &=&
\sum_{\mathbf{k}\in \mathcal{D}_j}
 \int \frac{d\omega}{2\pi} (\gamma_C|\mathcal{G}_{13}|^2 + \gamma_X  |\mathcal{G}_{14}|^2 )[\omega]
 \int \frac{d\omega'}{2 \pi}(\gamma_C|\mathcal{G}_{11}|^2 + \gamma_X
 |\mathcal{G}_{12}|^2 )[\omega'] \\
 \langle \hat{N}_s \hat{N}_i\rangle  - \langle \hat{N}_i\rangle\langle \hat{N}_s\rangle
 &=&\iint \frac{d\omega_1 d\omega_2 }{(2\pi)^2}\sum_{k\in \mathcal{D}_s}
\left( \gamma_C\mathcal{G}_{11}^*[\mathbf{k},2\omega_p
-\omega_1]\mathcal{G}_{13}^*[-\mathbf{k},\omega_1]+
\gamma_X\mathcal{G}_{12}^*[\mathbf{k},2\omega_p
-\omega_1]\mathcal{G}_{14}^*[-\mathbf{k},\omega_1]
\right) \nonumber \\
 &\phantom{=}&\left(\gamma_C\mathcal{G}_{11}[\mathbf{k},2\omega_p -\omega_2] \mathcal{G}_{13}[-\mathbf{k},\omega_2]
 +\gamma_X\mathcal{G}_{12}[\mathbf{k},2\omega_p -\omega_2] \mathcal{G}_{14}[-\mathbf{k},\omega_2]
\right)~.
\end{eqnarray}
\twocolumngrid

\begin{figure}[!ht]
\begin{center}
\includegraphics[width=9cm]{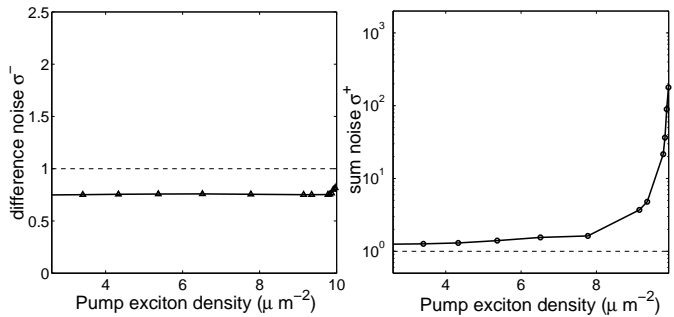}
\caption{\label{analytic_sum_diff} Analytical results for the
photonic difference and sum noises $\sigma^\pm$ as a function of
the pump exciton density $\rho_{p}$ in the absence of
disorder. Same cavity and integration parameters as in
Fig.\ref{cavity_photon_kx_ky_1}. $\hbar \omega_p = \hbar
\omega_{LP}(\mathbf{k}_p)+ 0.07$ meV $ = $ 1398 meV and $\hbar\omega_X^0 =
1400.1$ meV.}
\end{center}
\end{figure}

The results for $\sigma^{-}$ and $ \sigma^+$ are plotted in Fig.
\ref{analytic_sum_diff}. The qualitative similarities between
these results and the QMC ones (without disorder) of Fig
\ref{sum_difference} are apparent for pump intensities up to the parametric threshold.
Here, the linearized model breaks down, as it predicts a diverging signal/idler intensity.
Although the predictions for the threshold pump intensity differs from the QMC one by
 approximately twenty percent, still the analytic value in the low intensity
  limit $\sigma^- \simeq 0.75$ is well within the
(quite large) error bars of the QMC simulations without disorder\cite{foot_comment}.
In this limit, the analytical calculation, that neglects interactions between the
signal and idler modes, becomes indeed exact and provides a more precise
estimation than the QMC calculation.
As we have already mentioned, partition noise due to the half photon half
exciton nature of the polaritons is responsible for a significantly larger
 value of $\sigma^-$ than in standard $\chi^{(2)}$ parametric emitters~\cite{Laurat03}.
While in the QMC calculations the inclusion of static disorder has been
done straightforwardly, the simplified analytical model can not be
extended easily to the disorder case, and, most of all, it would
lose all its simplicity.

\section{Conclusion}
\label{Conclusions}

In conclusion, we have presented a quantum Monte Carlo study of
the quantum correlations in the ring-shaped parametric
luminescence from semiconductor microcavities in the strong
exciton-photon coupling regime. Our results suggest that even in
presence of multiple scattering, realistic losses and static
disorder, the signal and idler beams maintain a significant amount
of quantum correlations. The dependance of the intensity quantum
correlation on the pump intensity has been characterized across
the parametric instability threshold, showing the regime where the
non-classical features are maximized.

We thank  G. Bastard, J. Bloch, A. Bramati, C. Diederichs, E. Giacobino, J-Ph. Karr, C. Leyder, N. Regnault, M. Romanelli, Ph. Roussignol and J. Tignon for discussions.


\begin{thebibliography}{}
\bibitem{note} We have
verified that the results do not qualitatively change when
different specific potential configurations are used.
\bibitem{foot_comment} A straighforward application of squ
eezing theory shows that the exact 
value $\sigma^- = 0.75$ is recovered in the limit where the upper polariton 
can be neglected and the lower polariton is exactly half photon and half exciton.
\end{thebibliography}
\end{document}